\pdfoutput=1
\documentclass[a4paper]{jpconf} 

\usepackage{graphicx}
\usepackage{amssymb} 
\usepackage{amsthm} 

\begin{document}

\title{Dilepton Measurements at STAR}
\author{Frank Geurts (for the STAR Collaboration)}
\address{Rice University, Houston, TX 77005, USA}
\ead{geurts@rice.edu}

\begin{abstract} 
  In the study of hot and dense nuclear matter, created in
  relativistic heavy-ion collisions, dilepton measurements play an
  essential role. Leptons, when compared to hadrons, have only little
  interaction with the strongly interacting system. Thus, dileptons
  provide ideal penetrating probes that allow the study of such a
  system throughout its space-time evolution. In the low mass range
  ($M_{ll}<1.1$~GeV/$c^2$), the dominant source of dileptons originates
  from the decay of vector mesons which may see effects from chiral
  symmetry restoration. In the intermediate mass range
  ($1.1<M_{ll}<3.0 $~GeV/$c^2$), the main contributions to the mass
  spectrum are expected to originate from the thermal radiation of a
  quark-gluon plasma as well as the decays of charm mesons. In the
  high mass range ($M_{ll}>3.0$~GeV/$c^2$), dilepton measurements are
  expected to see contributions from primordial processes involving
  heavy quarks, and Drell-Yan production.

  With the introduction of the Time-of-Flight detector, the STAR
  detector has been able to perform large acceptance, high purity
  electron identification. In this contribution, we will present
  STAR's recent dielectron measurements in the low and intermediate
  mass range for RHIC beam energies ranging between 19.6 and 200~GeV.
  Compared to electrons, muon measurements have the advantage of
  reduced bremsstrahlung radiation in the surrounding detector
  materials.  With the upcoming detector upgrades, specifically the
  muon detector (MTD), STAR will be able to include such measurements
  in its (di-)lepton studies. We will discuss the future dilepton
  program at STAR and the physics cases for these upgrades.
\end{abstract}

\section{Introduction}
Dileptons have long been proposed as one of the more crucial probes in
the study of the hot and dense matter \cite{Shuryak78}. Leptons do not
experience the strong force, and thus will have negligible final state
interactions in a strongly interacting medium, with a mean free path
that is much larger than the typical size of this medium.
Electromagnetic probes such as the dielectrons are emitted throughout
the evolution of a heavy-ion collision. A typical dielectron
invariant mass spectrum, therefore, involves a plethora of sources,
ranging from Dalitz decays, which dominate the low invariant mass
range, to Drell-Yan pair production which dominates at masses above
3~GeV/$c^2$. In addition, dilepton contributions originate from vector
mesons, open heavy-flavor decays, and thermal radiation emitted from a
quark-gluon plasma (QGP).

The subsequent stages in the evolution of a hot and dense nuclear
system can be identified with certain ranges in the dilepton invariant
mass spectrum. In the high invariant mass range
($M_{ll}>3$~GeV/$c^2$), dileptons from the decay of the heavy
quarkonia such as the $J/\psi$ and $\Upsilon$ mesons provide a means
to study deconfinement effects in the hot and dense medium. These
contributions are on top of a continuum from primordial Drell-Yan pair
production. In the intermediate mass range ($1.1<M_{ll}<3.0
$~GeV/$c^2$), the production of dileptons is closely related to the
thermal radiation of the QGP. However, at higher center-of-mass
energies this signal competes with significant contributions from open
heavy-flavor decays such as $\mathrm{c}\bar{\mathrm{c}}\rightarrow
e^+e^-X$, where such charm contributions may get modified by the
medium. The prominent sources of dileptons in the low mass range
($M_{ll} <$ 1.1 GeV/$c^2$), in addition to the Dalitz decays, are the
direct leptonic decays of the $\rho(770)$, $\omega(782)$, and
$\phi(1020)$ vector mesons. The in-medium modification of the spectral
shape of these mesons may serve as signature of chiral symmetry
restoration. The $\rho$ meson is of special interest given that in
thermal equilibrium its contribution to the low mass range is expected
to dominate through its strong coupling to the $\pi\pi$ channel
\cite{RappWambachHees}. Moreover, its short lifetime of $\tau\sim
1.3$~fm/$c$, will make the spectral shape of this meson especially
sensitive to in-medium modifications \cite{CERES2}.

At SPS energies, the observed low-mass dilepton enhancement observed
in both the CERES dielectron \cite{CERES2,CERES1} and NA60 dimuon data
\cite{NA60} could be explained in terms of in-medium effects on the
spectral shape of the $\rho$ meson. Moreover, the dimuon measurements
by NA60 are found to favor significant broadening of the line shape
over a mass-dropping model.  At top RHIC energies, the PHENIX
collaboration measured a significant enhancement in its dielectron
measurements, with a strong $p_\mathrm{T}$ and centrality dependence
\cite{PHENIX}. However, models that have been able to describe the
measurements at SPS energies are unable to consistently describe the
PHENIX results in the most central collisions. The intermediate
invariant mass range, between the $\phi$ and $J/\psi$ mesons, opens an
important window to thermal radiation from the QGP.  However,
contributions from the earlier mentioned semileptonic decays of open
heavy-flavor hadrons are significant and need to be accounted for.

Direct photon measurements by the PHENIX collaboration in the range of
$1< p_\mathrm{T}< 4$~GeV/$c$ yielded a substantial elliptic
flow, $v_2$, comparable to the $v_2$ of hadrons \cite{PHENIXv2}. Model
calculations which include a dominant QGP thermal emission source
significantly underpredicted the observed $v_2$. However, incorporating a
more detailed evolution of the hadronic phase, which involves the
hadronic flow fields at chemical and kinetic freeze-out, appeared to
improve the description of the direct photon $v_2$ reasonably well,
while setting further constraints on the initial QGP temperatures
\cite{Hees}.  Dilepton elliptic flow measurements as a function of
$p_\mathrm{T}$ have been proposed as an independent measure to study
the medium properties \cite{Chatterjee}. The combination of certain
invariant mass and transverse momentum ranges allows for different
observational windows on specific stages of the expansion. Dileptons
can be used to further probe the early stages after a collision and
possibly constrain the QGP equation of state.

The installation of the Time-of-Flight (TOF) detector \cite{startof}
and the upgrade of its data acquisition (DAQ) system \cite{stardaq},
allowed the STAR experiment to extend its large-acceptance particle
identification capabilities and increase its DAQ rate. The TOF
detector not only extends the reach of hadron identification to higher
momenta, but also significantly improves the electron identification
capability in the low momentum range. This, combined with the RHIC
Beam Energy Scan (BES) program in 2010 and 2011, has put STAR in a
unique position to measure dielectron spectra in the low and
intermediate mass ranges from top RHIC beam energies down to SPS
center-of-mass energies. In this paper, the preliminary dielectron
results from the STAR experiment at top RHIC energy,
$\sqrt{s_\mathrm{NN}}=200$~GeV, are discussed as well as the
preliminary results at several BES energies. The low invariant mass
measurements are compared with recent model calculations. This paper
concludes with an outlook on the future of the STAR dilepton program.

\section{Electron Identification and Background Reconstruction}
The electron identification for the results reported in the next
sections involves the STAR Time Projection Chamber (TPC) and the TOF
detector. The TPC detector is the central tracking device of the STAR
experiment and it provides charged particle tracking and momentum
measurement. The energy-loss measurements, dE$/$d$x$, in the TPC are
used for particle identification. The TOF detector, with full
azimuthal coverage at mid-rapidity, extends the particle
identification range to higher momenta. The combination of the TOF
velocity information and the TPC energy loss allows for the removal of
slower hadrons which contaminate the electron sample. The selection
criteria for electron identification have been optimized for each RHIC
beam energy. With an additional track momentum threshold of
$p_\mathrm{T}>0.2$~GeV/$c$, the electron purity in the minimum bias
Au$+$Au analysis is 95\%.

\begin{figure}[ht]
\begin{minipage}[t]{0.45\linewidth}
\centering
\includegraphics[width=\textwidth,keepaspectratio]{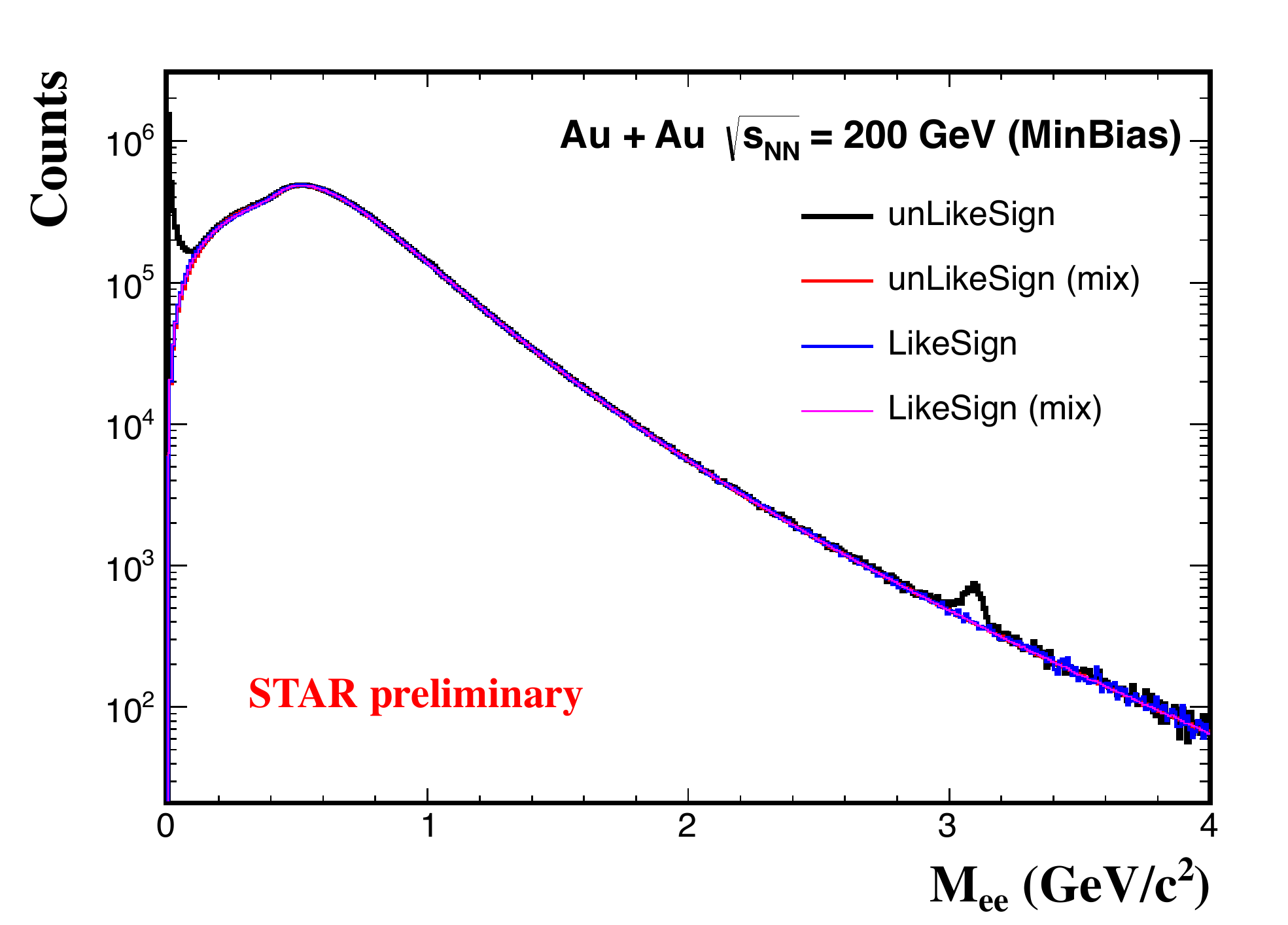}
\caption{(Color online) Unlike-sign, like-sign, and mixed-event
  dielectron invariant mass distributions in Au$+$Au minimum bias
  collisions at $\sqrt{s_\mathrm{NN}}=200$~GeV.}
\label{fig:backgroundAuAu}
\end{minipage}
\hspace{0.05\linewidth}
\begin{minipage}[t]{0.5\linewidth}
\centering
\includegraphics[width=\textwidth,keepaspectratio]{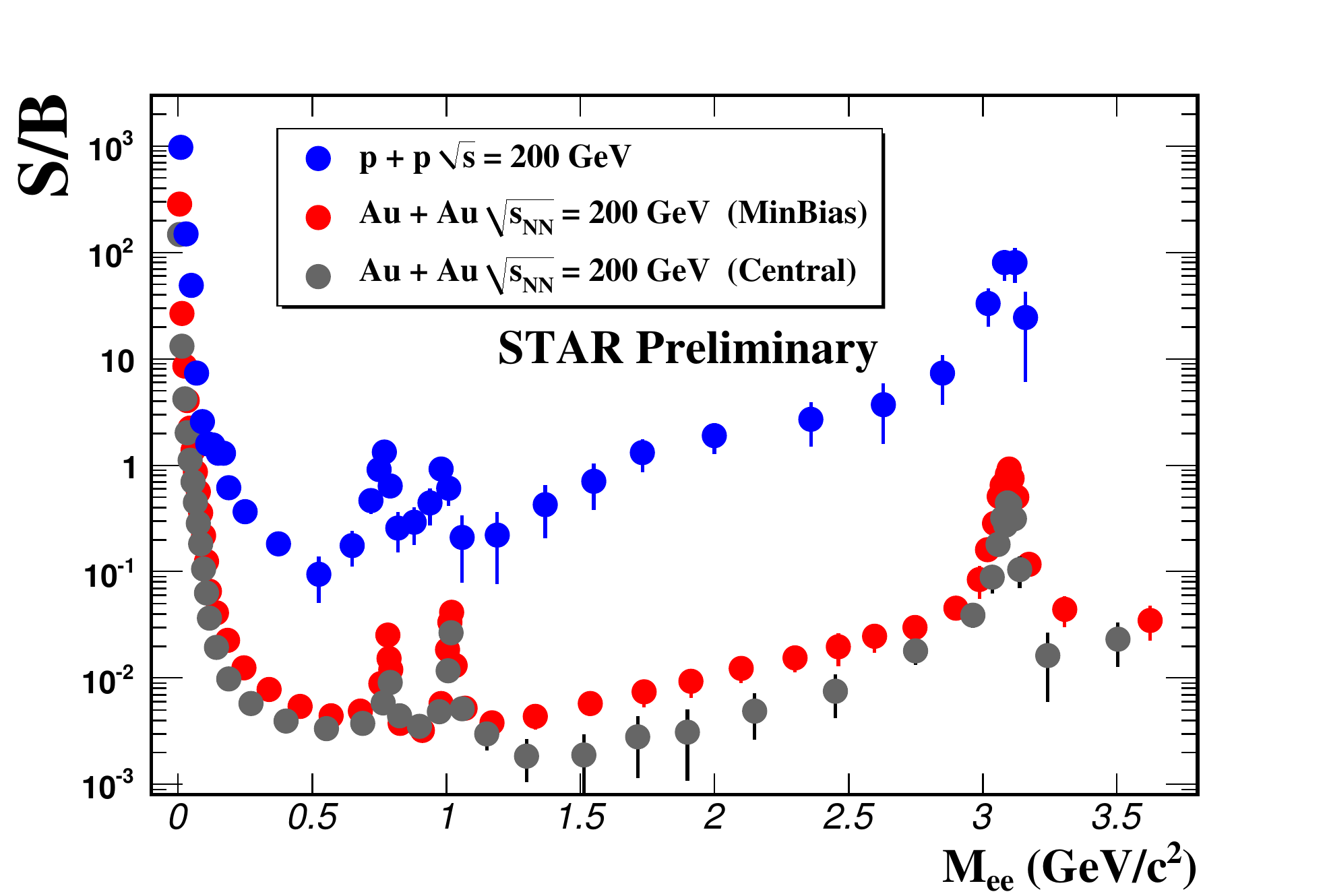}
\caption{(Color online) Signal-to-background ratios in p$+$p, and
  Au$+$Au central, minimum bias events at
  $\sqrt{s_\mathrm{NN}}=200$~GeV \protect\cite{ZhaoQM11}.}
\label{fig:sbratio}
\end{minipage}
\end{figure}

The unlike-sign invariant mass distributions, which are reconstructed
by combining electrons and positrons from the same event, contain both
signal and background contributions. Especially in high-multiplicity
events, the contribution of the combinatorial background is
large, see Fig.\ \ref{fig:backgroundAuAu}, and two methods have been
used to estimate the background.

The mixed-event method combines electrons and positrons from different
events with similar total particle multiplicity, vertex position along
the beam line, and event-plane angle. While the statistical accuracy
of the background description can be arbitrarily improved by involving
more events, the mixed-event method fails to reconstruct correlated
background sources.  At the lower invariant masses, such correlated
pairs arise from jets, double Dalitz decays, Dalitz decays followed by
a conversion of the decay photon, or two-photon decays followed by the
conversion of both photons \cite{Ruan2011}. A background estimation
based on the like-sign method can account for such correlated
contributions. Its drawback, however, is that the statistical accuracy
is only comparable to the unlike-sign, {\em i.e.}  the original raw
mass spectrum.  Moreover, the like-sign method will need to consider
detector acceptance differences, in contrast to the (unlike-sign)
mixed-event method.

The like-sign method is applied throughout the full mass range in
Au$+$Au at $\sqrt{s_\mathrm{NN}}=$ 19.6~GeV, in the low mass range for
$M_{ee}<750$~MeV/$c^2$ at $\sqrt{s_\mathrm{NN}}=$ 200~GeV, and
for $M_{ee}<900$~MeV/$c^2$ at 39~GeV and 62.4~GeV. Above these
respective invariant mass thresholds, the mixed-event method is
applied \cite{ZhaoQM11,HuangQM12}. The signal-to-background ratios for
200~GeV p$+$p and Au$+$Au are shown in Fig.\ \ref{fig:sbratio}.

\section{Dielectron Measurements at $\sqrt{s_\mathrm{NN}}=200$~GeV}
The STAR experiment has recently published its results on dielectron
measurements in p$+$p at $\sqrt{s}=$200~GeV \cite{starpp}. These
results were based on 107 million p$+$p events taken in the 2009 RHIC
run, with only a partially installed TOF system. The agreement between
the measured yields and the expected yields from a range of hadronic
decays, heavy-flavor decays, and Drell-Yan production, provided an
important verification of the analysis methods. More recently, the
STAR experiment significantly improved its p$+$p data sample by about
700 million events with a fully installed TOF, effectively doubling
its dielectron acceptance.

\begin{figure}[ht]
\centering
\includegraphics[width=0.9\textwidth,keepaspectratio]{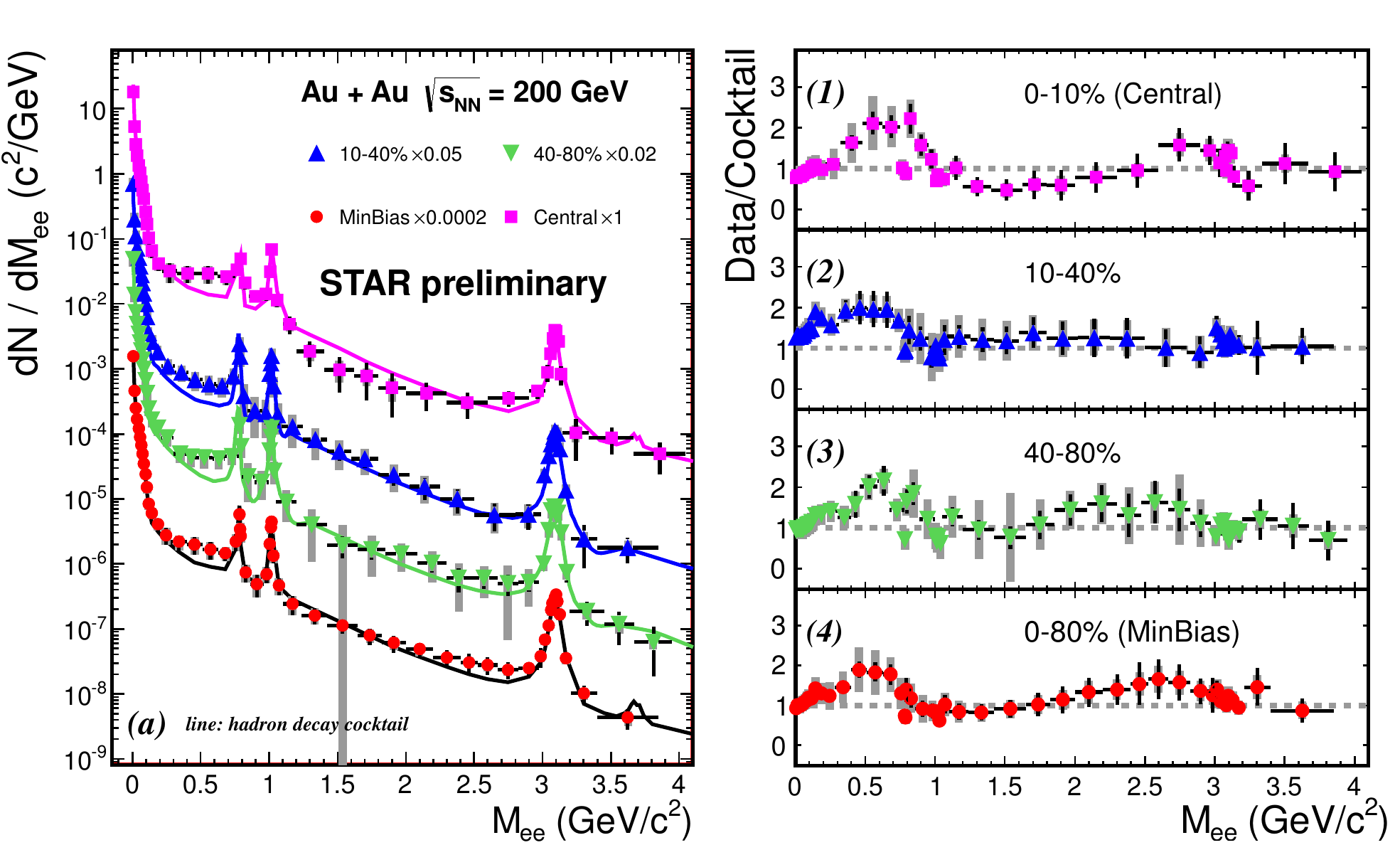}
\caption{(Color online) Dielectron invariant mass spectra for Au$+$Au collisions
 at
  $\sqrt{s_\mathrm{NN}}=$200~GeV for different centrality selections
  (left panel) and the ratio of data to cocktail (right
  panel). The systematical uncertainties are indicated by boxes.}
\label{fig:dielectronSpectra200GeV}
\end{figure}

The left panel of Fig.\ \ref{fig:dielectronSpectra200GeV} presents for
Au$+$Au collisions at $\sqrt{s_\mathrm{NN}}$=200~GeV the centrality
dependence of the invariant mass spectra in the STAR acceptance
($|y_{ee}|<1.0$, $|\eta_e|<1$, and $p_\mathrm{T}> 0.2$~GeV/$c$). The
measured yields are compared to a cocktail simulation of expected
yields where the hadronic decays include the leptonic decay channels
of the $\omega$, $\phi$, and $J/\psi$ vector mesons, as well as the
Dalitz decays of the $\pi^0$, $\eta$, $\eta'$ mesons \cite{ZhaoQM11}.
The input distributions to the simulations are based on Tsallis
Blast-Wave function fits to the invariant yields of the measured
mesons \cite{Ruan2011}. These functions serve as the input
distributions for the {\sc Geant} detector simulation using the full
STAR detector geometry. The $\rho$ meson contributions have not been
included in the cocktails as it may be sensitive to in-medium
modifications which are expected to affect this meson's spectral line
shape \cite{starrho}. In the intermediate mass range, the
c$\bar{\mathrm{c}}$ cross section is based on {\sc Pythia} simulations
scaled by the number of nucleon-nucleon collisions \cite{starCharm}.
The cocktail simulations are observed to overestimate the data in
central collisions. This can indicate a modification of the charm
contribution. However, the observed discrepancy is still consistent
within the experimental uncertainties.

\begin{figure}[ht]
\begin{minipage}[t]{0.48\linewidth}
\centering
\includegraphics[width=\textwidth,keepaspectratio]{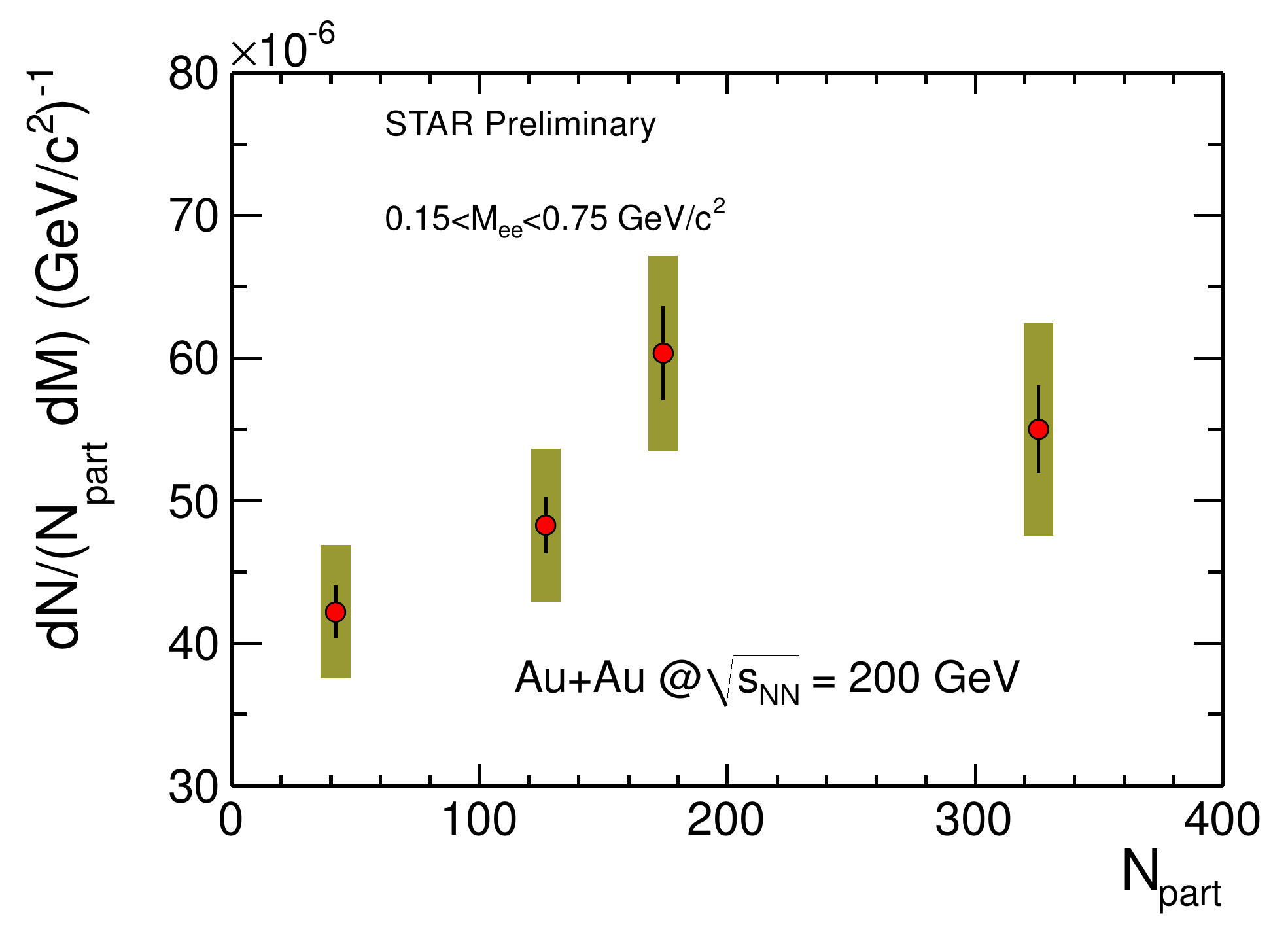}
\caption{(Color online) Dielectron LMR yields scaled by
  $N_\mathrm{part}$ versus centrality ($N_\mathrm{part}$) for Au$+$Au
  collisions at $\sqrt{s_\mathrm{NN}}$$=$200~GeV \cite{RuanQM12}. The
  boxes indicate the systematical uncertainty.}
\label{fig:LMRenhancement200GeV}
\end{minipage}
\hspace{0.06\linewidth}
\begin{minipage}[t]{0.46\linewidth}
\centering
\includegraphics[width=\textwidth,keepaspectratio]{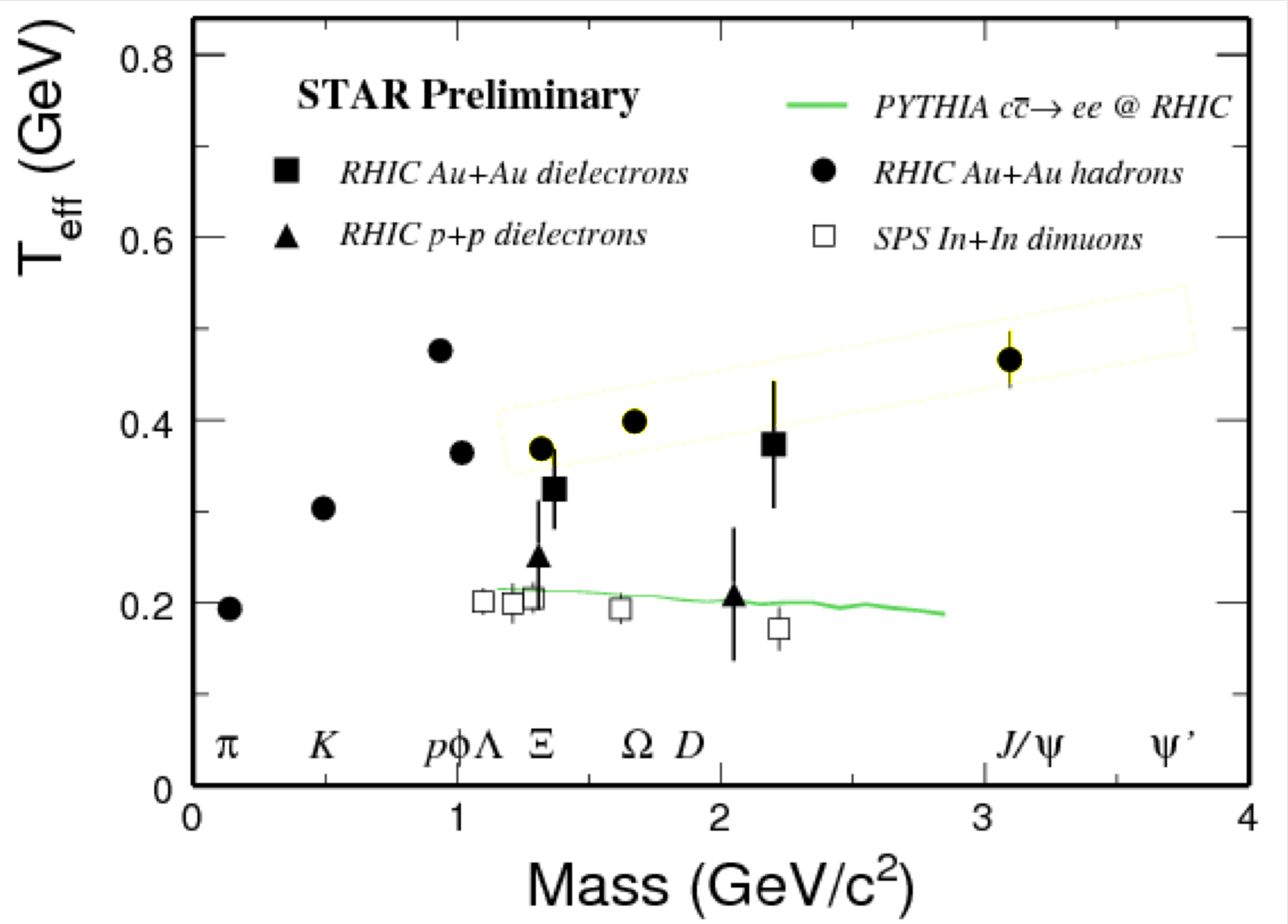}
\caption{Transverse mass slope parameters from Au$+$Au
  at $\sqrt{s_\mathrm{NN}}$$=$200~GeV (RHIC) and In$+$In at
  $\sqrt{s_\mathrm{NN}}$$=$17.2~GeV (SPS) energies \cite{ZhaoQM11}.}
\label{fig:IMRmT}
\end{minipage}
\end{figure}

In the right panels of Fig.\ \ref{fig:dielectronSpectra200GeV}, the
ratios of the data to cocktail yields have been depicted for different
centrality selections. A clear enhancement in the low mass range can
be observed. As the charm contributions scale with the number of
binary collisions, the total cocktail yield increases with centrality,
and only little centrality dependence can be observed in the ratio
plots. On the other hand, as can be seen in Fig.\
\ref{fig:LMRenhancement200GeV}, a comparison of the dilepton yield
$dN/dM_{ee}$ in the range of $150 < M_{ee} < 750$~MeV/$c^2$ scaled to
the number of participants, $N_\mathrm{part}$, appears to indicate an
increase of the low-mass-range enhancement with increasing centrality.
Such an increase, albeit with large uncertainties, would agree with
similar observations reported in \cite{CERES1}.

Figure \ref{fig:IMRmT} shows the inclusive dielectron transverse mass
slopes, in the intermediate mass range, measured in p$+$p and Au$+$Au
at $\sqrt{s_\mathrm{NN}}=$200~GeV. The $m_\mathrm{T}$ slopes in this
plot have been determined for $1.1<M_{ee}<1.8$,
$1.8<M_{ee}<2.8$~GeV/$c^2$ in Au$+$Au and $1.1<M_{ee}<1.6$,
$1.6<M_{ee}<2.9$~GeV/$c^2$, respectively. These measurements are
compared with the slope parameters of hadrons (circles) and
charm/Drell-Yan subtracted dimuon measurements (open squares,
\cite{NA60mT}). While the p$+$p results (triangles) are consistent
with {\sc Pythia} calculations, the transverse mass slopes in Au$+$Au
collisions (filled squares) are observed to be larger than those in
p$+$p.  This is indicative of a possible combination of both thermal
dilepton production and charm modification. Future detector upgrades
will allow STAR to further disentangle the potentially modified charm
contributions and help provide improved measurements of the thermal
QGP dilepton radiation.

\begin{figure}[ht]
\centering
\includegraphics[width=0.47\textwidth,keepaspectratio]{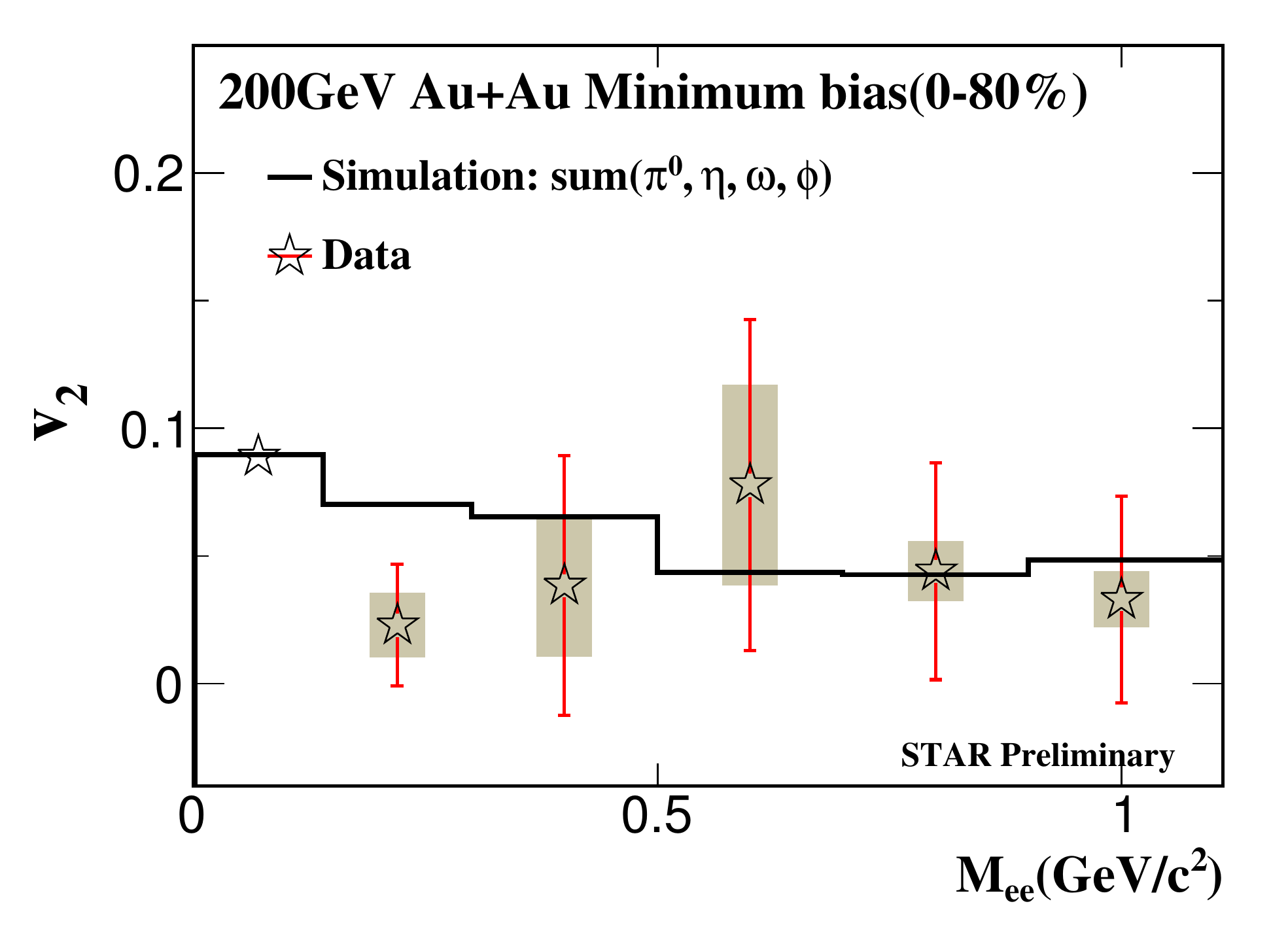}
\caption{(Color online) Dielectron elliptic flow as a function of the
  invariant mass in minimum-bias Au$+$Au collisions at
  $\sqrt{s_\mathrm{NN}}=$200~GeV. The boxes indicate the systematic
  uncertainty. The black line is the sum of a simulation which
  involved $\pi^0$, $\eta$, $\omega$, and $\phi$-mesons.}
\label{fig:dielectronflow}
\end{figure}

The STAR dielectron elliptic-flow measurements in Au$+$Au at
$\sqrt{s_\mathrm{NN}}=200$~GeV as a function of the dielectron
invariant mass are presented in Fig.\ \ref{fig:dielectronflow}.
These preliminary results are based on 700 million minimum bias events
involving an analysis that combined the data sets from the 2010 and
2011 RHIC Runs. The elliptic flow, $v_2$, is calculated using the
event-plane method in which the event plane has been reconstructed
from TPC tracks \cite{starflow}. The ``signal'' elliptic flow,
$v_2^\mathrm{signal}$, is calculated as follows \cite{HuangSQM11}:
\[
 v_2^\mathrm{total}(M_{ee}) =  v_2^\mathrm{signal}(M_{ee}) \frac{r(M_{ee})}{1+r(M_{ee})}  + v_2^\mathrm{bkgd}(M_{ee}) \frac{1}{1+r(M_{ee})}, 
\]
where $v_2^\mathrm{total}$ is the flow measurement of all dielectron
candidates, $v_2^\mathrm{bkgd}$ the flow measurement of the dielectron
background, and $r(M_{ee})$ the mass-dependent
signal-to-background ratio. The expected $v_2$ from a cocktail
simulation based on the contributions from $\pi^0$, $\eta$, $\omega$,
and $\phi$ mesons is within uncertainties consistent with the
measurements.

\begin{figure}[ht]
\centering
\includegraphics[width=0.6\textwidth,keepaspectratio]{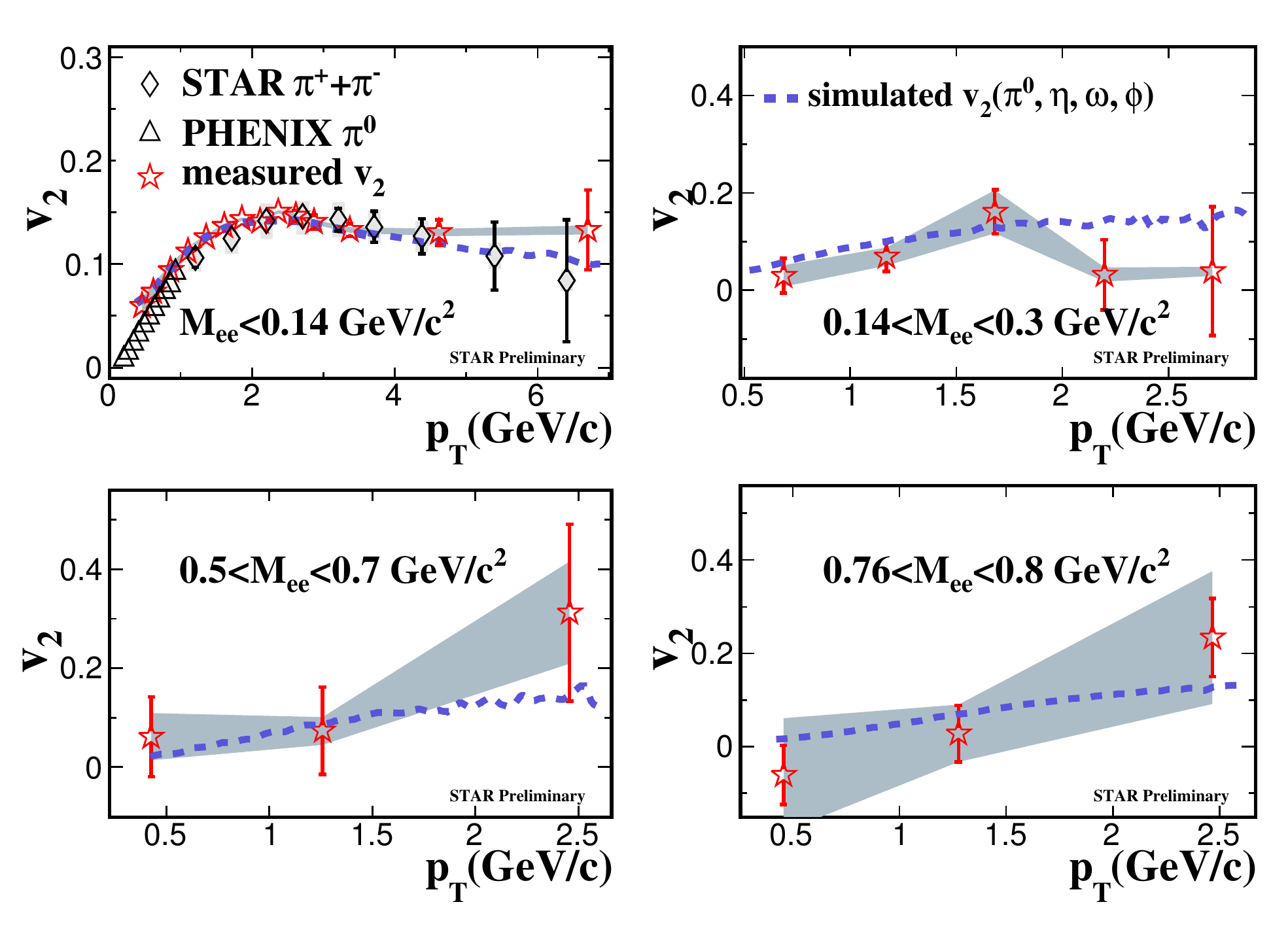}
\includegraphics[width=0.3\textwidth,keepaspectratio]{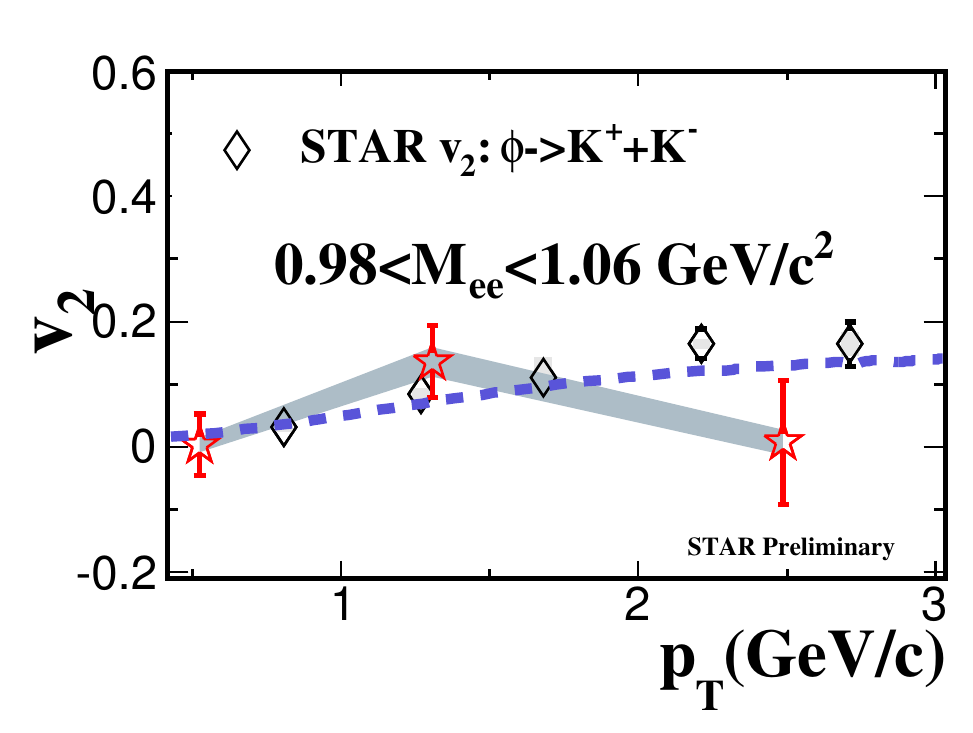}
\caption{(Color online) Elliptic flow, $v_2$, as a function of
  $p_\mathrm{T}$ for various dielectron invariant mass ranges in
  minimum-bias collision Au$+$Au collisions at
  $\sqrt{s_\mathrm{NN}}=$200~GeV (red stars). In addition, in the
  upper left panel the $v_2$ of $\pi^0$ \cite{phenixpi0} and $\pi^\pm$
  mesons \cite{starpiv2} are shown. The $\phi$ meson $v_2$
  measurements in the lower right panel are from STAR. The blue dashed
  lines are the expected $v_2$ from cocktail simulations (see text).
  The grey bands indicate the systematic uncertainties.}
\label{fig:flowpt}
\end{figure}

Differential measurements have been done as a function of centrality
(not shown here) and $p_\mathrm{T}$, as shown in Fig.\
\ref{fig:flowpt}. Both $v_2(p_\mathrm{T})$ for different dielectron
mass windows, and its centrality dependence in the lower mass bin show
a consistency between the measurements and the simulations. Work is
underway to further extend these measurements into the intermediate
mass range. However, it will be important to disentangle the charm
contributions at higher $M_{ee}$. The current experimental
uncertainties on the STAR data points are still too large to allow for
further constraints on the QGP equation of state, as is conjectured in
{\em e.g.} \cite{Chatterjee}.

\section{Dielectron Measurements in the Beam Energy Scan Program}
Measurements performed at top RHIC energies by both the STAR and PHENIX
\cite{PHENIX} collaborations indicate a significant enhancement in the
low mass range when compared to a hadron cocktail simulation of the
expected yields. Such an enhancement may point to in-medium
modification effects that possibly result from chiral symmetry
restoration. At the SPS, measurements performed by the NA60
\cite{NA60} and CERES \cite{CERES2, CERES1} collaborations also show
a low-mass enhancement which favors a broadening of the $\rho$ meson
spectral shape when compared to a dropping-mass scenario. The Beam
Energy Scan (BES) program at RHIC allowed the STAR collaboration to
systematically explore the dielectron production from
$\sqrt{s_\mathrm{NN}}=200$~GeV down to SPS beam energies.

\begin{figure}[ht]
\begin{minipage}[t]{0.55\linewidth}
\centering
\includegraphics[width=\textwidth,keepaspectratio]{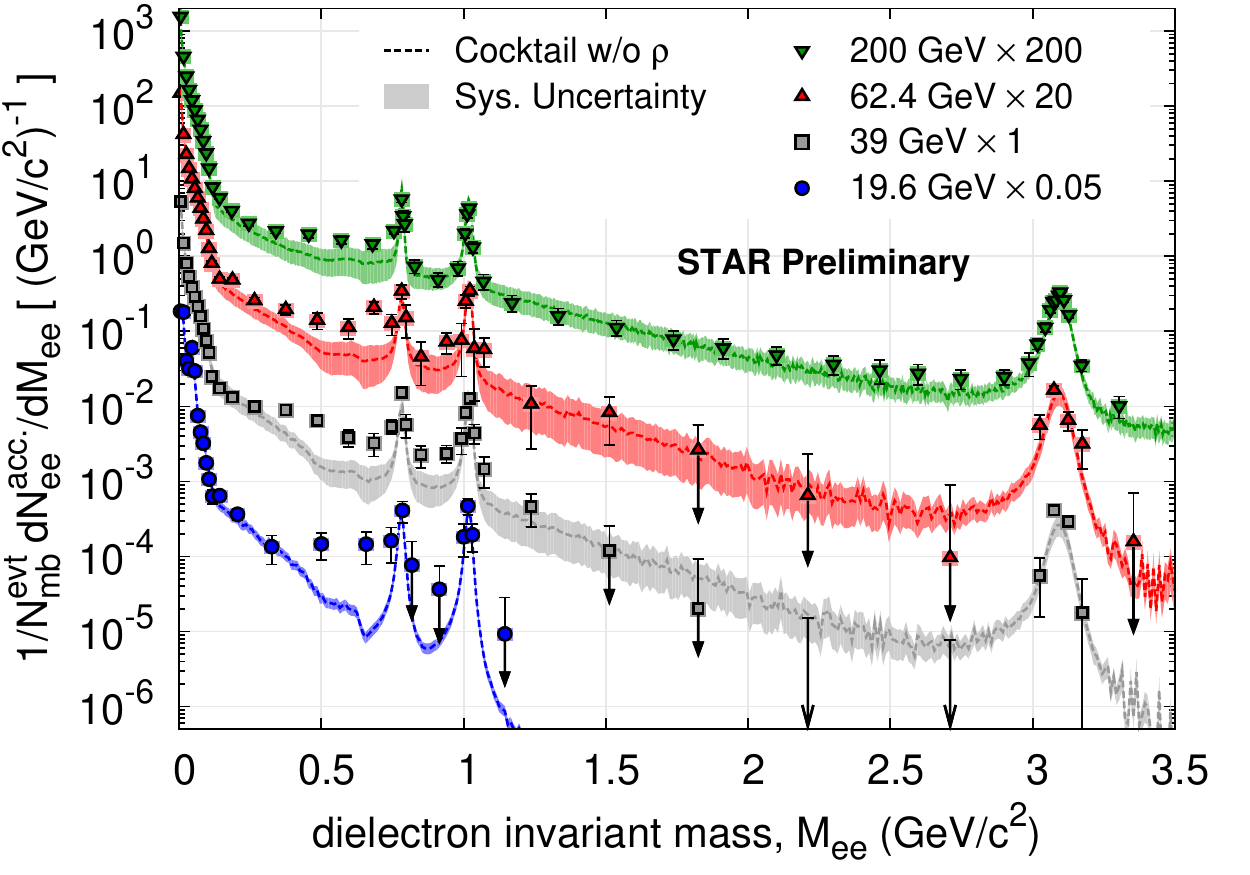}
\caption{(Color online) Background-subtracted dielectron
  invariant-mass distributions from Au$+$Au collisions at
  $\sqrt{s_\mathrm{NN}}=$ 19.6, 39, 62.4, and 200~GeV. The (colored)
  dotted lines show the hadron cocktails (excluding contributions from
  $\rho$ mesons). The (color) shaded areas indicate the systematic
  uncertainties.}
\label{fig:allenergiesdata}
\end{minipage}
\hspace{0.05\linewidth}
\begin{minipage}[t]{0.4\linewidth}
\centering
\includegraphics[width=\textwidth,keepaspectratio]{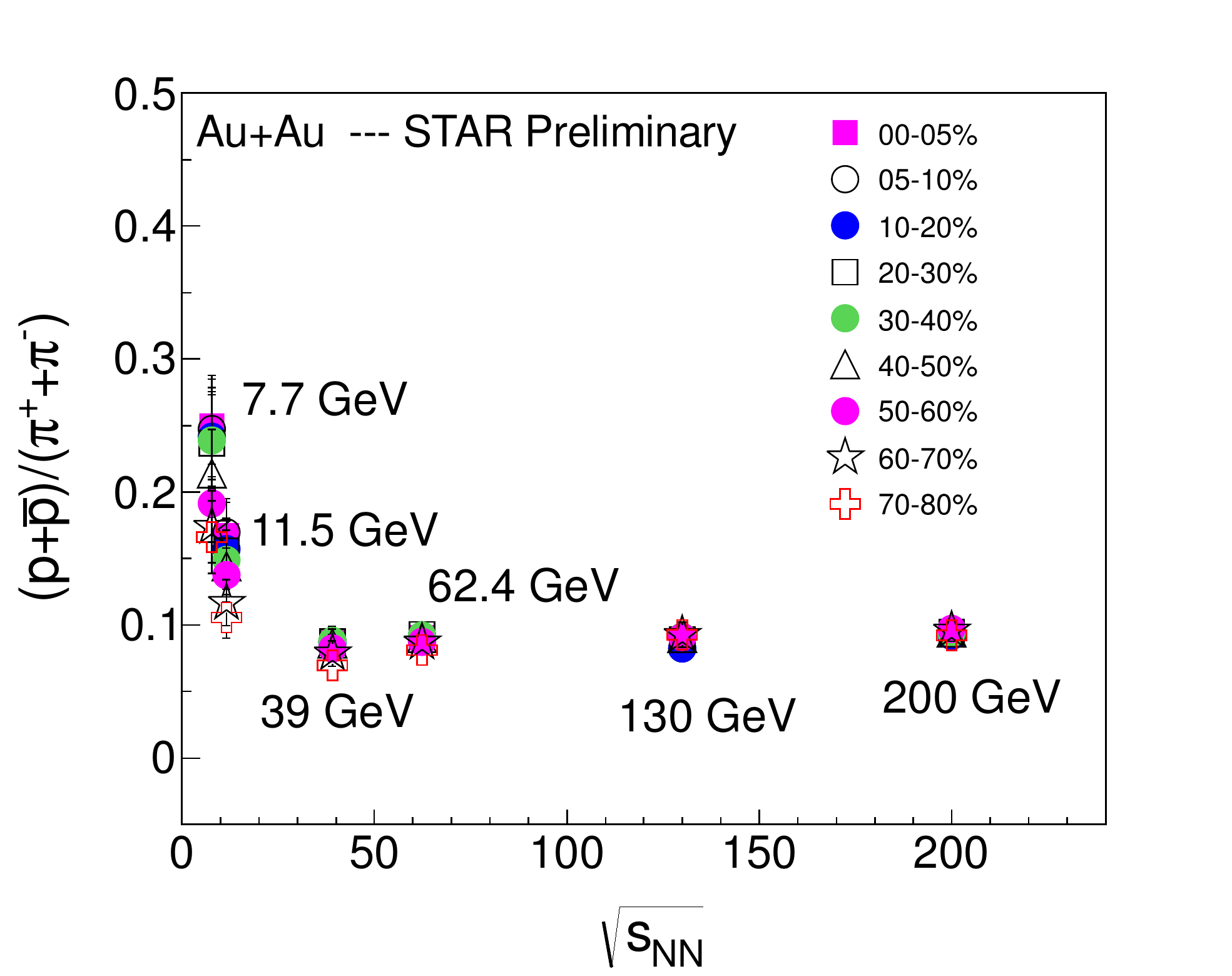}
\caption{(Color online) $(\mathrm{p}+\bar{\mathrm{p}})/(\pi^+ +
  \pi^-)$ in Au$+$Au collisions as a function of the center-of-mass
  energy $\sqrt{s_\mathrm{NN}}$.}
\label{fig:totalbaryon}
\end{minipage}
\end{figure}

In Fig.\ \ref{fig:allenergiesdata}, the inclusive invariant mass
spectra in Au$+$Au collisions for$\sqrt{s_\mathrm{NN}}=$ 19.6, 39,
62.4, and 200~GeV are shown, together with the hadron cocktail
simulations for each energy. The cocktail simulations exclude
contributions from the $\rho$ meson. For each of the energies a
significant enhancement in the low mass range can be observed. While a
quantitative discrepancy remains between the PHENIX and STAR
measurements at $\sqrt{s_\mathrm{NN}}=$ 200~GeV
\cite{PHENIX,geurtsAustin}, the preliminary STAR results are
comparable with the CERES measurements in the Pb$+$Au system at
$\sqrt{s_\mathrm{NN}}=$17.2~GeV albeit with differences in acceptance
\cite{CERES2,HuangQM12}. Moreover, various models have shown good
agreement with the low-mass spectrum measured by STAR in central
Au$+$Au collisions at $\sqrt{s_\mathrm{NN}}=200$~GeV
\cite{PHSD,USTC,Rapp}. The in-medium broadening of the $\rho$ meson is
expected to be driven by the strong coupling to baryons and thus the
total baryon density since the vector mesons interact symmetrically
with baryons and antibaryons \cite{CERES2,Rapp}. At SPS beam energies
with substantial nuclear stopping, most baryons are participating
nucleons. At RHIC top energies nuclear stopping and thus the
net-baryon density will vanish and a significant baryon-antibaryon
production is expected to compensate the total baryon density. As
shown in Fig.\ \ref{fig:totalbaryon}, the total baryon density at
freeze-out as a function of $\sqrt{s_\mathrm{NN}}$ does not change
significantly with center-of-mass energies down to ~20~GeV.
Accordingly, models that show good agreement at SPS and top-RHIC
energies should be able to describe the low-mass enhancement
throughout the BES center-of-mass energies.

\begin{figure}[ht]
\centering
\includegraphics[width=0.9\textwidth,keepaspectratio]{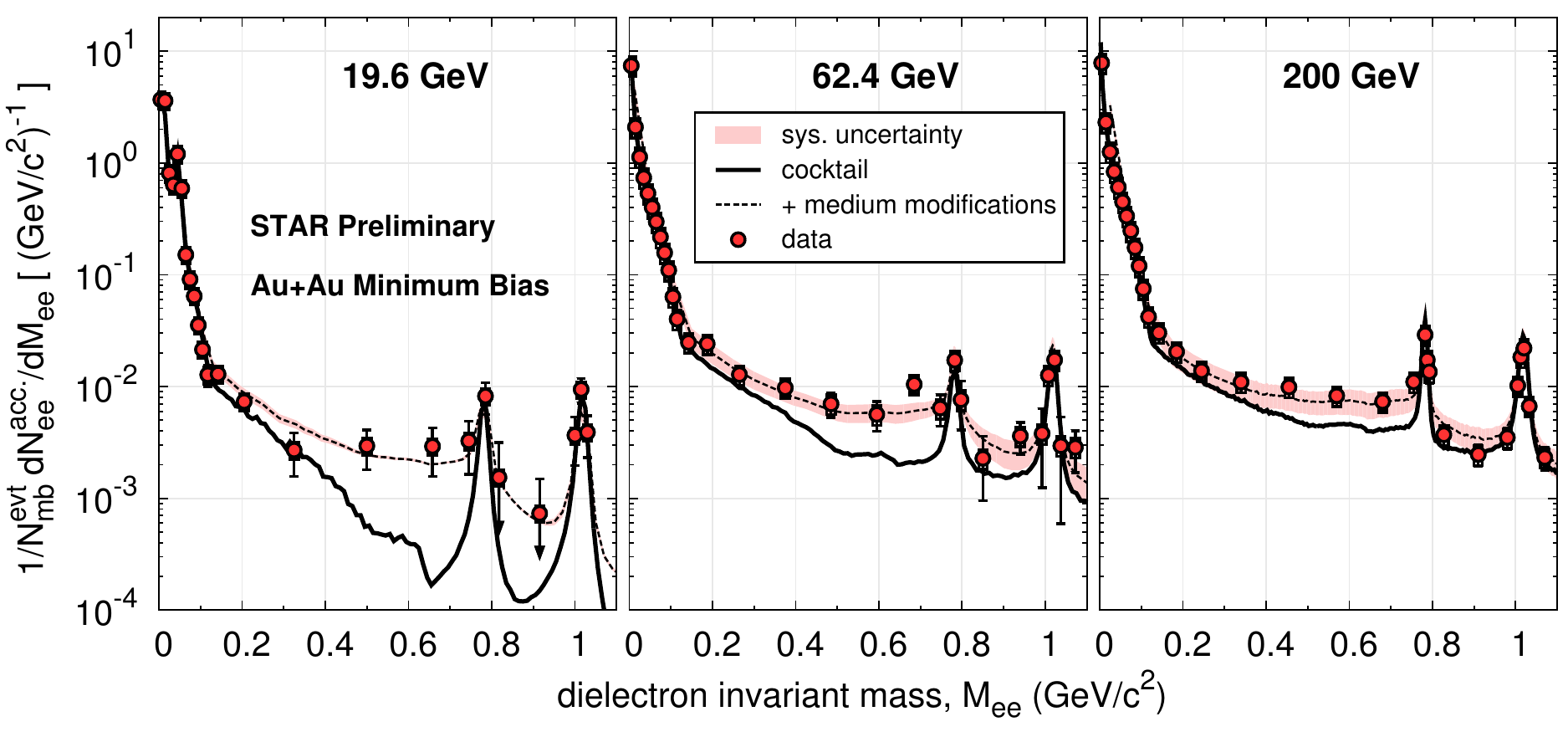}
\caption{(Color online) BES dielectron invariant-mass distributions in
  the low invariant-mass range from Au$+$Au collisions at 19.6~GeV
  (left), 62.4~GeV (middle), and 200~GeV (right panel). The red data
  points include both statistical and systematic uncertainties
  (boxes). The black curve depicts the hadron cocktail, while the
  dashed line shows the sum of the cocktail and model calculations.
  The latter includes contributions from both the HG and the QGP
  phases.  The systematic uncertainty on the former is shown by the
  light red band.}
\label{fig:inmediumrho}
\end{figure}

In Fig.\ \ref{fig:inmediumrho}, the efficiency-corrected invariant
mass spectra are shown for minimum bias Au$+$Au collisions at
$\sqrt{s_\mathrm{NN}}=$ 19.6, 62.4, and 200~GeV, respectively. In each
of the three panels the hadron cocktail simulations include
contributions from Dalitz decays, photon conversions (19.6~GeV only),
and the dielectron decay of the $\omega$ and $\phi$ vector mesons. As
is the case for the previously described 200~GeV cocktail simulations,
contributions from $\rho$ mesons have been excluded. Instead, the $\rho$
meson is explicitly included in the model calculations by Rapp
\cite{RappWambachHees,Rapp} which involve in-medium modifications of
the $\rho$ meson spectral shape. In this model a complete evolution of
the QGP and thermal dilepton rates in the QGP and hadron-gas (HG)
phases are convoluted with an isentropic fireball evolution. In the HG
phase the $\rho$ ``melts'' when extrapolated close the conjectured
phase transition boundary.  Moreover, it has been noted that the
top-down extrapolated QGP rates closely coincide with the bottom-up
extrapolated hadronic rates \cite{RappWambachHees}.
The measured low-mass enhancement can consistently be described by
these model calculations and agrees with a scenario in which the
in-medium modification of the $\rho$ involves a broadening of its
spectral function.

Differential studies of the low-mass dielectron distributions as
a function of the center-of-mass energy, centrality, and
$p_\mathrm{T}$ will allow for a more detailed comparison against these
chiral hadronic models. Further comparisons between these models,
lattice QCD calculations, and experimental data could help provide
explicit evidence of chiral symmetry restoration in heavy-ion
collisions.

\section{Future Directions of the STAR Dilepton Program}
Measuring dileptons is a challenging task where very small signals sit
on top of large combinatorial and physics backgrounds. With the recent
TOF upgrade, the STAR experiment is able to address some of the
important physics questions that involve the low and intermediate
dielectron mass ranges. In the low mass range, STAR's measurements
agree with results at SPS energies in which the enhancement is
attributed to the broadening of the $\rho$ meson. Moreover, model
calculations by Rapp \cite{Rapp} appear to be able to describe the BES
data at the intermediate center-of-mass energies. In the future,
additional verification of the robustness of such a description will
be allowed by the inclusion of independent analyses of other BES
data sets. Differential measurements will further verify the
consistency of such models and advance the study of chiral symmetry
restoration.

As the typical production rates for dileptons are rather small, large
event samples are needed to make significantly accurate measurements
in both the low and intermediate mass ranges. Especially at the lower
RHIC beam energies, the statistical uncertainties are very large (see
Fig.\ \ref{fig:allenergiesdata}).  The $\sqrt{s_\mathrm{NN}}$=19.6~GeV
measurements are based on 28 million Au$+$Au collisions. To achieve
statistical uncertainties at a level of ~10\% and improve the
understanding of the baryonic component to the in-medium effects on
the vector mesons, a factor of 10 more statistics would be required.
For such an increase in total luminosity at RHIC, requires the
development and installation of electron cooling components.

Cocktail simulations indicate that at top RHIC energies the correlated
charm contributions dominate the intermediate-mass range dielectrons.
The {\sc Pythia} simulations that are used for the cocktail
simulations assume no medium effects and have been rescaled to match
the charm cross sections to, {\em e.g.}, 0.96~mb in 200~GeV Au$+$Au
collisions. Both assumptions result in large uncertainties, with some
hints that point to a suppression of the charm contribution in central
collisions when compared to the minimum-bias data (see Fig.\
\ref{fig:dielectronSpectra200GeV}). These uncertainties directly
affect the extraction of any intermediate-mass dilepton signals
related to the thermal QGP radiation.  With the dominant source of
intermediate-mass $e-\mu$ correlations originating from the
c$\bar{\mathrm{c}}$ correlated pair decays, measuring this correlation
will be an essential tool for isolating the QGP thermal contribution
in the intermediate mass range. Upcoming STAR detector upgrades
\cite{VidebaekWWND13} will, among many other key measurements,
significantly improve our understanding of the charm contribution at
intermediate dilepton masses. The Muon Telescope Detector (MTD) will
provide a dedicated trigger for and the measurement of these
electron-muon correlations, where both leptons are measured at
mid-rapidity.

With the primary focus of the MTD physics program \cite{MTD} on the
quarkonia measurements at RHIC energies, the large-acceptance muon
detector will also allow the STAR dilepton physics program to include
the dimuon measurements of QGP thermal radiation, light vector mesons,
and Drell-Yan production. The MTD is expected to be fully commissioned
in 2014.

\ack{}
 This work is supported by the US Department of Energy under
grant DE-FG02-10ER41666.

\end{document}